\documentclass[12pt]{article}
\usepackage{graphicx,amssymb,amsmath,color}

\allowdisplaybreaks[4]
\numberwithin{equation}{section}

\newcommand{\ghs}{\gamma_{s}}

\newcommand{\dd}{\mathrm{d}}
\newcommand{\nn}{\nonumber}
\newcommand{\wides}{\widetilde{S}}
\newcommand{\SH}{\bar{S}^{\rm \,H}}

\newcommand{\gcusp}{\Gamma_{\rm cusp}}
\def\e{\epsilon}
\def\Ord{{\cal O}}
\def\Neqfour{{{\cal N}=4}}
\def\Nc{N_c}
\def\n3lo{N$^3$LO}
\def\dda{\mathcal{D}_0}
\def\ddb{\mathcal{D}_1}
\def\ddc{\mathcal{D}_2}
\def\ddd{\mathcal{D}_3}
\def\dde{\mathcal{D}_4}
\def\ddf{\mathcal{D}_5}
\def\ddg{\mathcal{D}_6}

\begin{document}

\catcode`\@=11
\font\manfnt=manfnt
\def\Watchout{\@ifnextchar [{\W@tchout}{\W@tchout[1]}}
\def\W@tchout[#1]{{\manfnt\@tempcnta#1\relax%
  \@whilenum\@tempcnta>\z@\do{%
    \char"7F\hskip 0.3em\advance\@tempcnta\m@ne}}}
\let\foo\W@tchout
\def\dubious{\@ifnextchar[{\@dubious}{\@dubious[1]}}
\let\enddubious\endlist
\def\@dubious[#1]{%
  \setbox\@tempboxa\hbox{\@W@tchout#1}
  \@tempdima\wd\@tempboxa
  \list{}{\leftmargin\@tempdima}\item[\hbox to 0pt{\hss\@W@tchout#1}]}
\def\@W@tchout#1{\W@tchout[#1]}
\catcode`\@=12

\thispagestyle{empty}

\null\vskip-20pt \hfill
\begin{minipage}[t]{72mm}
SLAC-PUB-16169, MITP/14-095\\
\end{minipage}
\vspace{5mm}

\begingroup\centering
{\Large\bfseries\mathversion{bold}
Soft-virtual corrections to Higgs production at N$^3$LO\par}%
\vspace{7mm}

\begingroup\scshape
Ye~Li$^{(1)}$, Andreas~von~Manteuffel$^{(2)}$,
\\Robert~M.~Schabinger$^{(2)}$, 
and Hua~Xing~Zhu$^{(1)}$\\
\endgroup
\vspace{5mm}
\begingroup\small
\emph{$^{(1)}$SLAC National Accelerator Laboratory\\
Stanford University, Stanford, CA 94309, USA}\\
\emph{$^{(2)}$PRISMA Cluster of Excellence \& Institute of Physics\\
Johannes Gutenberg University, 55099 Mainz, Germany}
\endgroup

\vspace{0.4cm}
\begingroup\small
E-mails:\\
{\tt yli@slac.stanford.edu}, {\tt manteuffel@uni-mainz.de},\\ {\tt rschabin@uni-mainz.de}, {\tt hxzhu@slac.stanford.edu}.
\endgroup
\vspace{0.7cm}

\textbf{Abstract}\vspace{5mm}\par
\begin{minipage}{14.7cm}
In this paper, we compute the soft-virtual corrections to Higgs boson production in gluon fusion for infinite top quark mass at next-to-next-to-next-to-leading order in QCD.
In addition, we present analogous soft-virtual terms for both Drell-Yan lepton production in QCD and scalar pair production in $\Neqfour$ super Yang-Mills theory.
The result for Drell-Yan lepton production is derived from the result for Higgs boson production using Casimir scaling arguments together with well-known results available in the literature.
For scalar pair production in the $\Neqfour$ model, we show by explicit calculation that the result is equal to the part of the Higgs boson soft-virtual term which is of maximal transcendentality weight.
\end{minipage}\par
\endgroup

\newpage



\section{Introduction}
\label{sec:intro}

The discovery of a scalar particle with properties compatible with the Standard Model Higgs boson was announced by the ATLAS and CMS collaborations on July 4, 2012~\cite{1207.7214,1207.7235}. 
After the discovery of this Higgs boson-like particle, detailed measurements of its properties became one of the top priorities of the ongoing experimental physics program at the Large Hadron Collider (LHC). 
To fully realize the physics potential of the machine in this regard, accurate and precise theoretical predictions are required for parton-initiated scattering processes involving the Standard Model Higgs boson. 

As has long been known, the primary mechanism for Standard Model Higgs boson production at the LHC is gluon-gluon fusion~\cite{PHLTA.40.692}. The leading order~(LO) cross section is of order $\alpha_s^2$ in Quantum Chromodynamics (QCD)
and is, unfortunately, neither accurate nor precise;
the next-to-leading order~(NLO) QCD corrections to this process in the limit of large top-quark mass were found to be about 100\%
of the LO result~\cite{NUPHA.B359.283,PHLTA.B264.440,hep-ph/9504378} and not covered by the conventional LO scale variation.
The importance of the gluon fusion process and the size of the NLO QCD corrections to it have motivated studies of the next-to-next-to-leading order (NNLO) QCD corrections, as well as
top quark mass effects and electroweak corrections. The next-to-next-to-leading order~(NNLO) QCD corrections to the total cross section in the large top-quark mass limit were computed in references~\cite{hep-ph/0201206,hep-ph/0207004,hep-ph/0302135}
and, in the same approximation scheme, the phenomenologically relevant, fully differential NNLO parton-level calculations were subsequently carried out in references~\cite{hep-ph/0501130,hep-ph/0703012,0707.2373,0801.3232}. 
The effects stemming from the finite mass of the top quark have been studied as well, both at NLO~\cite{Graudenz:1992pv} and at NNLO~\cite{0801.2544,0909.3420} in QCD. In addition, both
two-loop electroweak corrections~\cite{hep-ph/0404071,hep-ph/0407249,0809.3667} and three-loop mixed electroweak-QCD corrections~\cite{0811.3458} have been considered. 

Over the years, a number of soft-gluon or high-energy resummation techniques have been used to complement and consistently
improve the available fixed-order QCD results~\cite{hep-ph/9611272,hep-ph/0306211,hep-ph/0508265,hep-ph/0508284,hep-ph/0512249,hep-ph/0603041,hep-ph/0605068,0809.4283,1303.3590}.
Currently, the state-of-art is a NNLO+NNLL QCD prediction which, however, still suffers from a nearly $10\%$ scale uncertainty~\cite{1101.0593}. 
This is actually larger than the experimental error which can be achieved in the long run at LHC~\cite{1307.7135}
and it is therefore essential to go beyond the NNLO QCD approximation to further reduce the theoretical uncertainty.
Important steps in this direction have already been taken by a number of groups. For example, both the effective Higgs-gluon coupling~\cite{hep-ph/9708255,hep-ph/0512058,hep-ph/0512060}
and the QCD beta function~\cite{PHLTA.B93.429,hep-ph/9302208,hep-ph/9701390,hep-ph/0411261} have been known to four-loop accuracy for quite some time. 
Somewhat more recently, three-loop virtual amplitudes for $gg \to H$ were calculated by two different groups~\cite{0902.3519,1001.2887,1004.3653,1010.4478}.
In particular, all purely virtual ingredients for the calculation of $gg \to H$ at next-to-next-to-next-to-leading order~(\n3lo) in QCD are well-known and available in the literature. 

Of course, in order to obtain the physical cross section at \n3lo, one still needs to deal with real radiation and convolve parton-level results with the appropriate splitting amplitudes~\cite{1211.6559}.
In contrast to the purely virtual corrections discussed above, a complete calculation of the real radiative corrections is still missing.
Although the complete set of single-emission real-virtual corrections have now been calculated~\cite{1311.1425,1312.1296,1411.3586,1411.3587},
even more substantial progress has been made by working in the threshold approximation where all secondary real radiation is soft.
In fact, very recently, the next-to-threshold contribution to the cross section became available as well~\cite{1411.3584}.
Before the exact results became available, the single-emission, two-loop real-virtual corrections in the soft limit were computed by two different groups in references~\cite{1309.4391,1309.4393}. 
The triple-emission real corrections were calculated in the soft limit in reference~\cite{1302.4379},
and a subset of the exact master integrals required for these contributions were obtained in~\cite{1407.4049}.
The soft double-emission, one-loop real-virtual corrections were given implicitly in reference~\cite{1403.4616}
and more explicitly so in reference~\cite{1404.5839}. Most importantly, the full set of \n3lo soft plus virtual corrections (or, as we shall hereafter call them, soft-virtual corrections) were obtained in reference~\cite{1403.4616}. 

In this paper, we continue our study of the \n3lo soft-virtual corrections to the gluon-fusion Higgs production process. 
To that end, we calculate squares of time-ordered matrix elements of pairs of semi-infinite Wilson line operators with up to three massless partons in the final state. The object that we calculate to three-loop order is commonly referred to as a
soft function for a pair of Wilson line operators and frequently appears in the study of observables in soft-collinear effective theory~(SCET)~\cite{hep-ph/0005275,hep-ph/0011336,hep-ph/0109045,hep-ph/0206152}
(see also reference~\cite{1410.1892} for a review). 
In a precise sense, the soft function encodes the non-trivial part of the threshold limit of the QCD cross section,
coming from correlations between soft partons radiated off of the incoming hard gluons responsible for the primary production process.
Given that all purely virtual corrections are known, the calculation of the \n3lo soft-virtual corrections of interest amounts to the calculation of the \n3lo, three-loop soft function. In earlier work, we calculated
both the single-emission, two-loop real-virtual~\cite{1309.4391} and the double-emission, one-loop real-virtual~\cite{1404.5839} contributions to the three-loop soft function. In this article, we complete the calculation by treating both
the square of the single-emission, one-loop real-virtual corrections and the triple-emission real corrections. 
These single-emission contributions to the three-loop soft function can be straightforwardly obtained by using the soft-gluon current at one loop~\cite{hep-ph/9903516,hep-ph/0007142}.
The triple-emission contributions, on the other hand, are highly non-trivial and their calculation requires the development of specialized computational technology.

Ultimately, we managed to overcome all technical hurdles and complete the calculation of the three-loop soft function initiated by two of us some time ago in reference~\cite{1309.4391}.
After briefly reviewing our SCET-based formalism in Section \ref{sec:formalism}, we present the complete three-loop soft function for Higgs boson production at threshold in Section \ref{sec:softHiggs}.
Then, by appropriately combining our results with the well-known purely virtual contributions to the cross section, we obtain the soft-virtual term for Higgs boson production at \n3lo in Section \ref{sec:softvirt}. 
We compare our results to a recent calculation~\cite{1403.4616} and find full agreement. While our soft function calculation is performed with adjoint Wilson lines,
it is straightforward to convert the results obtained using Casimir scaling arguments to the case of fundamental Wilson lines appropriate for Drell-Yan lepton production
(see {\it e.g.}~\cite{0901.1091,0903.1126} for a discussion of the Casimir scaling principle).
By combining the resulting three-loop Drell-Yan soft function and the known three-loop virtual amplitudes for $\gamma^* \to q \bar{q}$~\cite{0902.3519,1001.2887,1004.3653,1010.4478},
we derive the soft-virtual term for Drell-Yan lepton pair production at \n3lo and present the result in Section \ref{sec:DY}. 
Once again, we find full agreement with recent independent predictions of the previously unknown part of the result~\cite{1404.0366,1405.4827}.

In fact, our highly-automated setup for the QCD squared eikonal matrix elements allows for an essentially trivial extension
to completely general massless final-state partons and, in particular, we can easily calculate the analogous three-loop soft function for a $SU(\Nc)$ $\Neqfour$ super Yang-Mills theory~\cite{Brink:1976bc}. 
In Section \ref{sec:SYM}, by making use of the $\Neqfour$ form factor computed in reference~\cite{1112.4524}, we give the \n3lo soft-virtual term for color-singlet scalar pair production in $\Neqfour$ super Yang-Mills theory.
Curiously, we find that this $\Neqfour$ soft-virtual term obeys a version of the principle of maximal transcendentality weight
first proposed long ago for the anomalous dimensions of twist-two operators in $\Neqfour$ super Yang-Mills theory~\cite{hep-th/0404092}. In this case, we find that our
result for the \n3lo soft-virtual term for scalar pair production in $\Neqfour$ super Yang-Mills theory coincides with the part of the \n3lo soft-virtual term for Higgs boson production which is of maximal transcendentality weight.
Finally, in Section \ref{sec:conclude}, we summarize our results and conclude the paper.


\section{Formalism}
\label{sec:formalism}

In this section, we briefly review the SCET formalism that we use in what follows.\footnote{We invite readers unfamiliar with SCET to consult reference~\cite{1410.1892} for a pedagogical introduction.} Although we will ultimately
be interested in other processes as well, our development in this section is specialized to the case of gluon-fusion Higgs boson production in order to keep the notation managable;
our discussion of the other processes treated in this paper will be completely analogous.
We are interested in the so-called threshold expansion of the partonic cross section, $\hat{\sigma}^{\mathrm{H}}_{gg} (\hat s, z, \alpha_s(\mu))$, in the limit $z\to 1$,
where $z = M^2_{\rm H}/{\hat s}$, $M_{\rm H}$ is the Higgs mass, and $\hat s$ is the square of the partonic center of mass energy.
In this regime, it can be written as
\begin{equation}
\label{eq:partxsect}
  \lim_{z\to 1}\left\{\hat{\sigma}^{\mathrm{H}}_{gg} (\hat s, z, \alpha_s(\mu))\right\} = \frac{\pi \lambda^2\left(\alpha_s(\mu),L_{\rm H}\right)}{ 8 (\Nc^2 - 1)} \, G^{\mathrm{H}}(z, L_{\rm H})\,,
\end{equation}
where $L_{\rm H} = \ln(\mu^2/M_{\rm H}^2)$ and $\lambda\left(\alpha_s(\mu),L_{\rm H}\right)$ is the effective coupling of the Higgs boson to gluons in the limit of infinite top quark mass,
$\mathcal{L}_{\mathrm{eff}} = - \frac{1}{4}\lambda H G^{\mu\nu,\,a}G^a_{\mu\nu}$. 
The factor out front of $G^{\mathrm{H}}(z, L_{\rm H})$ in Eq. (\ref{eq:partxsect}) is the LO cross section in this limit.
For the sake of convenience, we use a common factorization and renormalization scale $\mu_F = \mu_R = \mu$.

The coefficient function $G^{\rm H}(z, L_{\rm H})$ defined in this way can be shown to factorize in the framework of SCET as $z\to 1$:
\begin{align}
G^{\rm H}(z, L_{\rm H}) = H^{\rm H} (L_{\rm H}) \bar{S}^{\rm \,H}(z, L_{\rm H})\,,
\label{eq:coef}
\end{align}
where the renormalized hard function, $H^{\rm H} (L_{\rm H})$, captures all purely virtual effects and the renormalized soft function, $\bar{S}^{\rm \,H}(z, L_{\rm H})$,
captures all effects coming from soft final-state gluon radiation very close to threshold.
Note that Eq.~(\ref{eq:coef}) depends implicitly on the renormalized strong coupling constant, $\alpha_s(\mu)$.

As explained in reference~\cite{1404.5839}, at order $n$ in QCD perturbation theory, the $n$-th perturbative coefficient of the hard function in SCET for Higgs production
can be extracted from the $n$-loop gluon form factor~\cite{0902.3519,1001.2887,1004.3653,1010.4478}. In fact, the relevant formulae are written down explicitly in reference \cite{1004.3653}, Eqs. (7.6), (7.7), and (7.9).
The hard coefficients that we need for our analysis can be derived by taking the complex square of Eq. (7.3) in that
work, setting $\mu = M_{\rm H}$, and then expanding in the strong coupling to third order. 
The perturbative coefficients of the soft function, on the other hand, require a dedicated calculation and, in fact, we present the third-order coefficient explicitly for the first time in Section \ref{sec:softHiggs}.

The bare soft function for Higgs production can be written as the square of a time-ordered matrix element of a pair of semi-infinite Wilson line operators,
\begin{eqnarray}
\label{eq:softdef}
 S^{\rm H}\left(z, L_{\rm H}\right) = \frac{M_{\rm H}}{\Nc^2 - 1} \sum_{\mbox{\tiny $X_s$ }}
\langle 0 | T\Big\{Y_{n} Y_{\bar{n}}^\dagger\Big\} \delta\Big(\lambda - \hat{\bf P}^0\Big) | X_s\rangle\langle X_s | T\Big\{ Y_{\bar{n}} Y_{n}^\dagger \Big\} | 0 \rangle\,.
\end{eqnarray}
As discussed in reference~\cite{1404.5839}, the soft function defined in Eq. (\ref{eq:softdef}) depends on $z$ and the bare strong coupling constant $\alpha_s$.
In this notation, the ratio of twice the energy of the soft QCD radiation to $M_{\rm H}$ is approximately $1-z$.
The summation in Eq. (\ref{eq:softdef}) is over all possible soft parton final
states, $|X_s\rangle$. The operator $\hat{\bf P}^0$ acts on the final state $|X_s\rangle$ according to
\begin{equation}
 \hat{\bf P}^0|X_s\rangle = 2 E_{X_s} |X_s\rangle\,,
\end{equation}
where $E_{X_s}$ is the energy of the soft radiation in final state $|X_s\rangle$. The Wilson line operators, $Y_{n}$ and $Y_{\bar{n}}^\dagger$, are respectively defined as in-coming, path-ordered $\left(\mathbf{P}\right)$
 and anti-path-ordered $\left(\overline{\mathbf{P}}\right)$ exponentials \cite{hep-ph/0412110},
\begin{eqnarray}
  Y_{n}(x) &=& \mathbf{P} \exp \left( i g \int^0_{-\infty} \dd s\, n \cdot A(n s+x) \right)
\nn
\\
  Y_{\bar{n}}^\dagger(x) &=& \overline{\mathbf{P}} \exp \left( -i g \int_{-\infty}^0  \dd s\, \bar{n} \cdot A
    (\bar{n}s+x) \right).
\end{eqnarray}
In the above, $A_\mu = A_{\mu}^a T^a$, where the $T^a$ are adjoint $\mathfrak{su}(N_c)$ matrices.
As usual, $n$ and $\bar{n}$ are light-like vectors whose space-like components are back-to-back and determine the
beam axis. For a generic four-vector, $k^\mu$, we have
\begin{equation}
n \cdot k = k^+ \qquad {\rm and} \qquad \bar{n} \cdot k = k^-
\end{equation}
given the usual definitions $k^+ = k^0 + k^3$ and $k^- = k^0 - k^3$.

The soft function obeys a Renormalization Group~(RG) equation which involves a convolution in momentum space. 
The convolution makes it somewhat inconvenient to work in momentum space directly and it is customary to take a Laplace transform to turn the convolution into a product,
\begin{align}
\label{eq:Laplace}
  \wides^{\rm \,H}( L_\omega ) = \int^1_{-\infty} \! \dd z \, \exp
  \bigg( - \frac{ M_{\rm H} ( 1-z )}{e^{\gamma_E} \omega }\bigg) \SH (z, L_{\rm H})\,.
\end{align}
In Eq. (\ref{eq:Laplace}), $\omega$ is the variable Laplace-conjugate to $z$, $L_\omega =\ln \left(\mu^2/\omega^2\right)$, and $\gamma_E$ is the Euler-Mascheroni constant. 
The perturbative coefficients of the Laplace-transformed soft function are simple polynomial functions of $L_\omega$ of degree $2 n$ at $n$-loop order.
In this case, the inverse Laplace transform can be evaluated in closed form:
\begin{align}
\label{eq:inverseLaplace}
  \SH(z, L_{\rm H})  = \lim_{\eta \to 0} \bigg\{\wides^{\rm \,H}\left(-\partial_\eta\right) \frac{1}{(1-z)^{1 - 2 \eta} }
  \left( \frac{M^2_{\rm H}}{\mu^2} \right)^\eta \frac{ \exp ( -2 \eta \gamma_E) }{\Gamma ( 2 \eta ) }\bigg\}\,.
\end{align}
It is therefore possible to perform the entire analysis in Laplace space and then, at the very end, appropriately apply Eq. (\ref{eq:inverseLaplace}) to recover the momentum space formulas of interest.

In Laplace space, the RG equation for the soft function is~\cite{hep-ph/9808389}
\begin{align}
  \frac{\dd \wides^{\rm \,H}(L_\omega)}{\dd \ln (\mu)} = \bigg( 2  \gcusp^{\rm H}(a) L_\omega  - 2 \ghs^{\rm H}(a) \bigg) \wides^{\rm \,H}(L_\omega)\,,
\label{eq:RG}
\end{align}
where $\gcusp(a)$ and $\ghs(a)$ are, respectively, the cusp anomalous
dimension and the soft anomalous dimension. They admit perturbative
expansions of the form
\begin{align}
\gcusp^{\rm H} (a) = \sum_{n=1}^\infty a^{n} \Gamma^{\rm H}_{n - 1} \qquad {\rm and} \qquad \ghs^{\rm H} (a) =  \sum_{n=1}^\infty a^{n} \gamma^{\rm H}_{n - 1} \,,
\label{eq:gdef}
\end{align}
where we have defined 
\begin{align}
 a = \frac{\alpha_s(\mu)}{4 \pi}\,. 
\end{align}
Their perturbative coefficients up to and including terms of $\Ord\left(a^3\right)$ are collected in the appendix. 
Eq.~(\ref{eq:RG}) can be solved straightforwardly order-by-order in $a$. 
Through to $\Ord(a^3)$, one can write down the results immediately by simultaneously making the replacements $L \to -L_\omega$, $\Gamma_i \to - \Gamma_i^{\rm H}$, and $\gamma_i^{\rm H} \to - \gamma_i^{\rm H}$
in the right-hand side of Eq. (55) of reference~\cite{0803.0342},
\begin{align}
\label{eq:expand}
 \wides^{\rm \,H}\left(L_\omega\right)  = & \,1 + a \bigg[ \frac{1}{2} \Gamma_0^{\rm H} L^2_\omega - \gamma_0^{\rm H} L_\omega + c^{\rm H}_1 \bigg]
+ a^2 \bigg[ \frac{1}{8} \Big(\Gamma_0^{\rm H}\Big)^2 L^4_\omega + 
\left( \frac{1}{6} \beta_0 \Gamma_0^{\rm H}  - \frac{1}{2} \gamma_0^{\rm H} \Gamma_0^{\rm H}  \right) L^3_\omega
\\
& +
\left( \frac{1}{2}\Gamma_1^{\rm H} + \frac{1}{2}c^{\rm H}_1\Gamma_0^{\rm H} - \frac{1}{2} \beta_0 \gamma_0^{\rm H} + \frac{1}{2} \Big( \gamma_0^{\rm H} \Big)^2\right) L^2_\omega
+ \Big( \beta_0 c^{\rm H}_1 - c^{\rm H}_1 \gamma_0^{\rm H} - \gamma_1^{\rm H} \Big) L_\omega + c^{\rm H}_2\bigg] 
\nn
\\
&
+ a^3 \bigg[ \frac{1}{48} \Big(\Gamma_0^{\rm H}\Big)^3 L_\omega^6 +
\left(\frac{1}{12} \beta_0 \Big( \Gamma_0^{\rm H}\Big)^2 - \frac{1}{8} \gamma_0^{\rm H} \Big( \Gamma_0^{\rm H}\Big)^2 \right) L_\omega^5 
+  \left( \frac{1}{4} \Gamma_0^{\rm H} \Gamma_1^{\rm H} +  \frac{1}{8} c^{\rm H}_1 \Big( \Gamma_0^{\rm H} \Big)^2  \right.
\nn
\\
&
\left.+ \frac{1}{12} \Big(\beta_0\Big)^2 \Gamma_0^{\rm H} - \frac{5}{12} \beta_0 \gamma_0^{\rm H} \Gamma_0^{\rm H} + \frac{1}{4} \Big(\gamma_0^{\rm H}\Big)^2 \Gamma_0^{\rm H}\right) L_\omega^4 
+ \left( \frac{1}{3} \beta_0 \Gamma_1^{\rm H} - \frac{1}{2}\gamma_0^{\rm H} \Gamma_1^{\rm H}  + \frac{1}{6}\beta_1\Gamma_0^{\rm H} \right.
\nn
\\
&
\left. + \frac{2}{3}\beta_0 c^{\rm H}_1\Gamma_0^{\rm H} - \frac{1}{2}c^{\rm H}_1\gamma_0^{\rm H} \Gamma_0^{\rm H} - \frac{1}{2}\gamma_1^{\rm H} \Gamma_0^{\rm H} - \frac{1}{3}\Big(\beta_0\Big)^2 \gamma_0^{\rm H} 
+ \frac{1}{2} \beta_0 \Big(\gamma_0^{\rm H}\Big)^2 - \frac{1}{6} \Big(\gamma_0^{\rm H}\Big)^3 \right) L_\omega^3 
\nn
\\
&
+
\left( \frac{1}{2} \Gamma_2^{\rm H} + \frac{1}{2} c^{\rm H}_1 \Gamma_1^{\rm H}  + \frac{1}{2} c^{\rm H}_2 \Gamma_0^{\rm H}   + \Big(\beta_0\Big)^2 c^{\rm H}_1
- \frac{1}{2} \beta_1 \gamma_0^{\rm H} - \frac{3}{2}  \beta_0 c^{\rm H}_1 \gamma_0^{\rm H}+ \frac{1}{2} c^{\rm H}_1 \Big( \gamma_0^{\rm H}\Big)^2\right.
\nn
\\
&
\left. - \beta_0 \gamma_1^{\rm H} + \gamma_0^{\rm H} \gamma_1^{\rm H} \vphantom{\frac{1}{2} \Gamma_2^{\rm H}}\right) L_\omega^2  
+ \Big( \beta_1 c^{\rm H}_1 + 2 \beta_0 c^{\rm H}_2 - c^{\rm H}_2 \gamma_0^{\rm H} - c^{\rm H}_1 \gamma_1^{\rm H} - \gamma_2^{\rm H} \Big) L_\omega + c^{\rm H}_3  \bigg]+\Ord\left(a^4\right)\,.
\nn
\end{align}

While $\beta_0$ and $\beta_1$ are nothing but the one- and two-loop coefficients of the QCD beta function (see Appendix \ref{sec:append}), the $c^{\rm H}_n$ in Eq. (\ref{eq:expand})
are {\it a priori} unknown matching coefficients which must be determined by an explicit computation at each order in QCD perturbation theory. 
With the understanding that $c^{\rm H}_0 = 1$, we can write
\begin{equation}
c_s^{\rm H}(a) = \sum_{n=0}^\infty a^n c^{\rm H}_n
\end{equation}
in analogy with Eqs. (\ref{eq:gdef}) above. 
The one- and two-loop matching coefficients, $c^{\rm H}_1$ and $c^{\rm H}_2$, were calculated for a soft function built out of
fundamental representation Wilson line operators long ago~\cite{hep-ph/9808389} and, as was pointed out in reference~\cite{0809.4283}, the results can be
converted to the adjoint representation case relevant to Higgs production using a simple Casimir scaling argument. For a theory with $N_f$ light quark flavors, they read
\begin{align}
c^{\rm H}_1 = & \,2 \zeta_2 C_A
\nn
\\
c^{\rm H}_2 = & \,\frac{1}{2!} \Big(c^{\rm H}_1\Big)^2 + \Delta c^{\rm H}_2 \,,
\end{align}
where
\begin{align}
  \Delta c^{\rm H}_2 = \left( \frac{2428}{81} + \frac{67\zeta_2}{9} - \frac{22\zeta_3 }{9} - 30 \zeta_4 \right)C^2_A  + \left( -\frac{328}{81} - \frac{10\zeta_2}{9} + \frac{4\zeta_3}{9} \right) C_A N_f
\end{align}
with $C_A = \Nc$ and $C_F = (\Nc^2 - 1)/(2\Nc)$.

Note that, in the above, we have explicitly factored out the terms that are fixed by the non-Abelian exponentiation theorem~\cite{PHLTA.B133.90,NUPHA.B246.231}. 
This formulation works well for our purposes since, at least through three-loop order, $c_s^{\rm H}(a)$ is naturally written as an exponential (see Section \ref{sec:DY}):
\begin{align}
  c_s^{\rm H}(a) = \exp \Big( a c^{\rm H}_1 + a^2 \Delta c^{\rm H}_2 + a^3 \Delta c^{\rm H}_3 + \cdots \Big) \, .
\label{eq:csexp}
\end{align}
One of the main goals of the present paper is to calculate the new contribution, $\Delta c^{\rm H}_3$, to the three-loop matching coefficient and this is the subject of the next section.

\section{The Higgs soft function at \n3lo}
\label{sec:softHiggs}

\begin{figure}[!t]
\begin{center}
  \includegraphics[width=.7\textwidth]{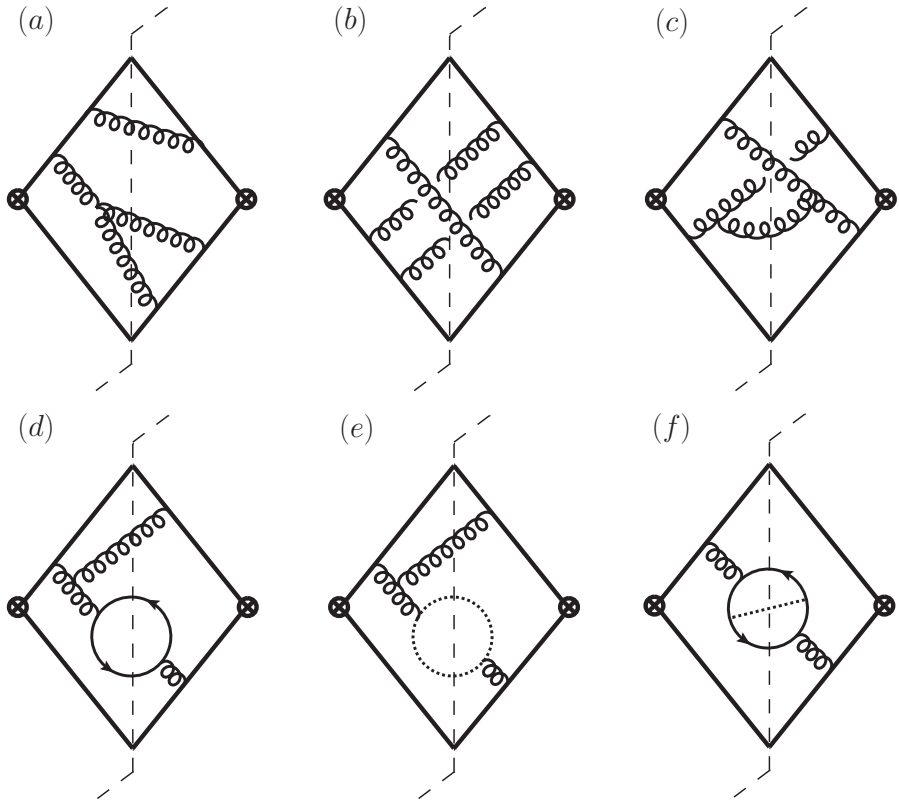}
  \caption{
  Cut eikonal Feynman diagrams for final states with $(a)-(c)$ three gluons, $(d)$ one gluon and two fermions, $(e)$ one gluon and two scalars, and $(f)$ one scalar and two fermions. 
  Diagrams such as the one shown in panel $(f)$ are only relevant for the calculation of the three-loop soft function in $\Neqfour$ super Yang-Mills theory. 
  Note that diagram $(b)$ is planar since it can be drawn in the plane without any lines crossing.
  Soft contributions from non-planar diagrams such as $(c)$ vanish for all of the processes considered in this paper because of the color Jacobi identity.}
\label{diag}
\end{center}
\end{figure}
In order to compute the threshold soft function for gluon-fusion Higgs boson production at \n3lo, one must consider the following:
\begin{enumerate}
\item Interference between tree-level and two-loop single soft emission diagrams.
\item Interference between tree-level and one-loop double soft emission diagrams.
\item The square of the one-loop single soft emission diagrams.
\item The square of the tree-level triple soft emission diagrams.
\end{enumerate}
As mentioned in the introduction, we computed the first two contributions in earlier work~\cite{1309.4391,1404.5839} and
the third contribution is straightforward to derive using the one-loop soft gluon current~\cite{hep-ph/9903516,hep-ph/0007142}. 
The fourth and final contribution is more challenging and we therefore briefly describe its calculation below. 

First, we generate all relevant tree-level triple soft emission diagrams using
{\tt QGRAF}~\cite{Nogueira:1991ex}. All possible partonic final states are considered, including $ggg$, $gq\bar{q}$, $g \phi\phi^\dagger$, and $q\bar{q}\phi$, where $\phi$ is an adjoint scalar. 
Representative cut eikonal diagrams are depicted in Figure~\ref{diag}.
To process the squared matrix element, we employ in-house {\tt FORM}~\cite{math-ph/0010025} and {\tt Maple} routines to carry out all of the necessary numerator algebra. 
The spacetime dimension is set to $D = 4 - 2 \e$ and the polarization dimension is set to $D_s$;  in conventional dimensional regularization~(CDR), $D_s = D$, and in the four-dimensional helicity scheme~(FDH),
$D_s=4$~\cite{hep-ph/0202271}.\footnote{The number of gluonic polarization states is $2 - 2 \e$ in the CDR scheme, and $2$ in the FDH scheme.}
As observed in our calculation of the interference between the tree-level and one-loop double soft emission diagrams~\cite{1404.5839}, it is essential to partial fraction the raw integrand. 
Once all independent topologies are identified, we apply the integration by parts reduction~\cite{PHLTA.B100.65,NUPHA.B192.159,hep-ph/0102033} algorithms for phase space integrals implemented in
 {\tt Reduze\;2}~\cite{1201.4330,cs.sc/0004015,fermat} and {\tt LiteRed}~\cite{1212.2685}. 
After reduction, a compact linear combination of seven master integrals is obtained. Our independent calculation of these integrals benefited from the development of a
new computational technique for phase space integrals which will be described in a forthcoming paper by one of us~\cite{zhu:2014}. 

Combining all of the contributions enumerated above, together with appropriate coupling constant and operator renormalization contributions\footnote{See {\it e.g.} Section II of reference~\cite{1408.5134} for a discussion
of the appropriate multiplicative renormalization.}, we obtain the renormalized soft function for
gluon-fusion Higgs boson production at \n3lo. An important and immediate cross-check is that our explicit calculation is completely consistent with the prediction of renormalization group invariance, Eq. (\ref{eq:expand}).
The three-loop matching coefficient, $c^{\rm H}_3$, is most naturally written in the exponentiated form (see Eq. (\ref{eq:csexp}))
\begin{align}
  c^{\rm H}_3 = & \frac{1}{3!}\Big(c^{\rm H}_1\Big)^3 + c^{\rm H}_1 \Delta c^{\rm H}_2 + \Delta c^{\rm H}_3\,,
\label{eq:expc3}
\end{align}
where, in the CDR scheme defined above, the last term in Eq.~(\ref{eq:expc3}),  $\Delta c^{\rm H}_3$  is given by
\begin{align}
  \Delta c^{\rm H}_3 = & \left(\frac{5211949}{13122} - \frac{20371\zeta_2}{729} - \frac{87052\zeta_3}{243} - \frac{9527\zeta_4}{27} - \frac{220\zeta_2\zeta_3}{9} - \frac{968\zeta_5}{9} \right.
\nn
\\
&
\left. + \frac{1072\zeta_3^2}{9} + \frac{8506\zeta_6}{27} \right)C^3_A + \left( -\frac{412765}{6561} +\frac{2638\zeta_2}{729} +\frac{1216\zeta_3}{81} +\frac{928\zeta_4}{27} \right.
\nn
\\
&
\left.- \frac{8\zeta_2\zeta_3}{9} -\frac{16\zeta_5}{3} \right) C_A^2 N_f + \left( -\frac{42727}{486} - \frac{55\zeta_2}{9} + \frac{2840\zeta_3}{81} + \frac{152\zeta_4}{9}\right.
\nn
\\
&
\left. + \frac{16\zeta_2\zeta_3}{3} + \frac{224\zeta_5}{9} \right) C_A C_F N_f + \left(-\frac{256}{6561} - \frac{8\zeta_2}{81} + \frac{880\zeta_3}{243} + \frac{52\zeta_4}{27}\right)C_A N_f^2\,.
\label{eq:delc3}
\end{align}
Eq.~(\ref{eq:delc3}) is one of the main results of this paper. 

\section{The soft-virtual term for Higgs production at \n3lo}
\label{sec:softvirt}

We now have all the ingredients required to write down the \n3lo soft-virtual corrections to gluon-fusion Higgs boson production. 
Combining Eqs.~(\ref{eq:coef}) and~(\ref{eq:inverseLaplace}), the coefficient function reads
\begin{align}
\label{eq:GHiggs}
G^{\rm H}(z, L_{\rm H}) = H^{\rm H} (L_{\rm H})  ~\lim_{\eta \to 0} \bigg\{\wides^{\rm \,H}\left(-\partial_\eta\right) \frac{1}{(1-z)^{1 - 2 \eta} }
                          \left( \frac{M^2_{\rm H}}{\mu^2} \right)^\eta \frac{ \exp (-2 \eta \gamma_E) }{\Gamma ( 2 \eta ) }\bigg\}\,.
\end{align}
For simplicity, we present the result for $\mu = M_{\rm H}$, such that $L_{\rm H} = 0$. Introducing the convenient short-hand notation for the distributions which appear in the soft-virtual term,
\begin{align}
\label{eq:dists}
\mathcal{D}_0 = \delta( 1 - z) \qquad {\rm and} \qquad \mathcal{D}_i = \left[ \frac{\ln^{i-1}( 1 - z)}{1 - z}\right]_+ \qquad \forall \, i > 0 \, ,
\end{align}
we have
\begin{align}
G^{\rm H}(z, 0) = & \dda + a \bigg\{ \bigg[ 16 \ddc + 8 \zeta_2 \dda\bigg] C_A \bigg\} 
+ a^2 \bigg\{ \bigg[ 128 \dde - \frac{176}{3} \ddd +  \left( \frac{1072}{9} - 160 \zeta_2 \right) \ddc
\nn
\\
&
 +  \left( -\frac{1616}{27} + \frac{176\zeta_2}{3}  + 312 \zeta_3 \right) \ddb + \left( 93 + \frac{536\zeta_2}{9}  - \frac{220\zeta_3}{3}  - 2 \zeta_4 \right) \dda \bigg]  C_A^2
\nn
\\
&
+ \bigg[ \frac{32}{3} \ddd - \frac{160}{9} \ddc + \left( \frac{224}{27}  - \frac{32\zeta_2}{3} \right) \ddb +  \left( - \frac{80}{3} - \frac{80\zeta_2}{9}  - \frac{8\zeta_3}{3} \right) \dda \bigg] C_A N_f
\nn
\\
&
+ \bigg[ \left( -\frac{67}{3} + 16 \zeta_3  \right) \dda \bigg] C_F N_f \bigg\} + a^3 \bigg\{ \bigg[ 512 \ddg - \frac{7040}{9} \ddf + \left( \frac{59200}{27} - 3584 \zeta_2 \right) \dde
\nn
\\
&
 +  \left( -\frac{67264}{27} + \frac{11968}{3} \zeta_2 + 11584 \zeta_3 \right) \ddd + \left(\frac{244552}{81} - \frac{9728\zeta_2}{3} - \frac{22528\zeta_3}{3}  \right.
\nn
\\
&
\left. - 4928\zeta_4 \vphantom{-\frac{67264}{27}}\right) \ddc +  \left( -\frac{594058}{729}+ \frac{137008\zeta_2}{81}  + \frac{143056\zeta_3}{27}  + \frac{2024\zeta_4}{3}  - \frac{23200\zeta_2 \zeta_3}{3}  \right.
\nn
\\
&
\left. + 11904 \zeta_5 \vphantom{-\frac{594058}{729}}\right) \ddb +  \left(\frac{215131}{81} + \frac{64604\zeta_2}{81}  - 3276\zeta_3 - \frac{30514\zeta_4}{27}  + \frac{7832\zeta_2 \zeta_3 }{3} - \frac{30316\zeta_5}{9}  \right.
\nn
\\
&
\left. + \frac{13216\zeta_3^2 }{3} - \frac{8012\zeta_6}{3} \vphantom{-\frac{594058}{729}}\right)  \dda \bigg]C_A^3 +  \bigg[ \frac{1280}{9} \ddf - \frac{10496}{27} \dde +  \left(\frac{14624}{27} - \frac{2176\zeta_2}{3} \right) \ddd 
\nn
\\
&
+  \left( -\frac{67376}{81} +\frac{6016\zeta_2}{9} + \frac{2944\zeta_3}{3}\right) \ddc + \left( \frac{125252}{729} - \frac{34768\zeta_2}{81}  - \frac{7600\zeta_3}{9} - \frac{272\zeta_4}{3}\right) \ddb
\nn
\\
&
 + \left(-\frac{98059}{81} -\frac{4480\zeta_2}{81}  + \frac{9848\zeta_3}{27} + \frac{10516\zeta_4}{27}- \frac{2000\zeta_2\zeta_3}{3} + \frac{6952\zeta_5}{9}\right) \dda \bigg] C_A^2 N_f
\nn
\\
&
 + \bigg[ 32 \ddd +  \Big( -504 + 384 \zeta_3 \Big) \ddc +  \left(\frac{3422}{27} - 32\zeta_2 - \frac{608\zeta_3}{9} - 32 \zeta_4 \right) \ddb
\nn
\\
&
 +  \left( -\frac{63991}{81} - \frac{1136\zeta_2}{9} + 416 \zeta_3 + \frac{88\zeta_4}{9}  + 192\zeta_2 \zeta_3 + 160\zeta_5 \right) \dda \bigg] C_A C_F N_f
\nn
\\
&
  +  \bigg[\left( \frac{608}{9} + \frac{592\zeta_3}{3}  - 320 \zeta_5 \right)  \dda\bigg]  C_F^2 N_f + \bigg[ \frac{256}{27} \dde - \frac{640}{27} \ddd +  \left( \frac{1600}{81} - \frac{256\zeta_2}{9}  \right) \ddc   
\nn
\\
&
 +  \left( -\frac{3712}{729}+ \frac{640\zeta_2}{27}  + \frac{320\zeta_3}{27}  \right)\ddb +  \left( \frac{2515}{27} - \frac{2128\zeta_2}{81}  + \frac{688\zeta_3 }{27} -\frac{304\zeta_4}{9} \right) \dda \bigg]C_A N_f^2 
\nn
\\
&
+  \bigg[\left( \frac{8962}{81} -\frac{184\zeta_2}{9} - \frac{224\zeta_3 }{3} - \frac{16\zeta_4}{9}   \right) \dda\bigg] C_F N_f^2 \bigg\} \, .
\label{eq:Hsv}
\end{align}
Remarkably, our calculation agrees completely both with the tower of plus distributions~\cite{hep-ph/0508265,hep-ph/0508284,hep-ph/0512249,hep-ph/0603041,hep-ph/0605068,0809.4283} and with the delta function terms computed for 
the first time in reference~\cite{1403.4616}. It should be stressed that, at every stage of the calculation, different computer codes and computational techniques were used by the authors of~\cite{1403.4616}.
Therefore, our analysis constitutes a highly non-trivial and important confirmation of their result.

\section{The soft-virtual term for Drell-Yan production at \n3lo}
\label{sec:DY}

It turns out that, once the three-loop soft function for gluon-fusion Higgs boson production is known, it
is not difficult to obtain the \n3lo soft-virtual corrections for Drell-Yan lepton production. 
First of all, the Drell-Yan hard function is known to $\Ord\left(a^3\right)$ and one can read off its one-, two-, and three-loop perturbative coefficients from  Eqs. (7.3), (7.4), (7.5), and (7.8) of reference~\cite{1004.3653}.
The one- and two-loop perturbative coefficients of the Drell-Yan soft function were computed long ago in reference~\cite{hep-ph/9808389} and, therefore, 
the three-loop soft function for the Drell-Yan process is the only non-trivial ingredient that remains.
It was observed in the QCD literature some time ago that the soft limit of the Drell-Yan production process can be
determined from the analogous result for Higgs boson production~\cite{hep-ph/0512249,hep-ph/0603041}.
Here, we present the analysis in the framework of SCET.

The first three perturbative coefficients of the Drell-Yan soft function can be obtained from the analogous results for Higgs boson production by making the replacements
\begin{gather}
  M_{\rm H} \to  M_{\gamma^*}
\qquad\qquad
\gcusp^{\rm H} (a) \to \frac{C_F}{C_A} \gcusp^{\rm H} (a)
\nn
\\
\ghs^{\rm H} (a) \to  \frac{C_F}{C_A} \ghs^{\rm H} (a)
\qquad\qquad
c_s^{\rm H}(a) \to  \left(c_s^{\rm H}(a)\right)^{\frac{C_F}{C_A}}
\label{eq:replacement}
\end{gather}
everywhere in the results for Higgs production presented in Eq.~(\ref{eq:expand}).\footnote{It is not known whether these replacement rules continue to hold at higher orders in QCD perturbation theory.}
In the above, $M_{\gamma^*}$ is the invariant mass of the off-shell photon intermediate state in the classical Drell-Yan process.
It becomes clear now why it is useful to define $c_s^{\rm H}(a)$ as in Eq.~(\ref{eq:csexp}):
although the soft matching coefficients $c_n^{\rm H}$ and $c_n^{\rm DY}$ need not be directly related,
subtracting exponentiated lower-order contributions reveals a simple Casimir scaling relation
between the remaining pieces, $\Delta c_n^{\rm H}$ and $\Delta c_n^{\rm DY}$, at least for $n \leq 3$.

In analogy to Eq. (\ref{eq:GHiggs}), we have
\begin{align}
  G^{\rm DY}(z, L_{\rm DY}) = H^{\rm DY} (L_{\rm DY})  ~\lim_{\eta \to 0} \bigg\{\wides^{\rm \,DY}\left(-\partial_\eta\right) \frac{1}{(1-z)^{1 - 2 \eta} }
                          \left( \frac{M_{\gamma^*}^2}{\mu^2} \right)^\eta \frac{ \exp ( -2 \eta \gamma_E) }{\Gamma ( 2 \eta ) }\bigg\}\,,
\end{align}
where $L_{\rm DY} = \ln(\mu^2/M_{\gamma^*}^2)$.
Going through the steps described above, we obtain the \n3lo soft-virtual corrections for Drell-Yan lepton production. 
For simplicity, we present the result for $\mu = M_{\gamma^*}$, such that $L_{\rm DY} = 0$. 
Explicitly, we have
\begin{align}
  G^{\rm DY}(z, 0) = & \dda + a \bigg\{ \bigg[ 16 \ddc + \Big( -16 + 8 \zeta_2 \Big) \dda\bigg] C_F \bigg\} + a^2 \bigg\{ \bigg[ 128 \dde + \Big( -256 - 128\zeta_2\Big) \ddc  
\nn \\ &
+ 256 \zeta_3 \ddb + \left( \frac{511}{4} - 70 \zeta_2  - 60 \zeta_3  + 4\zeta_4 \right) \dda \bigg]  C_F^2 +  \bigg[ - \frac{176}{3} \ddd +  \left( \frac{1072}{9} \right. 
\nn \\ &
\left. - 32 \zeta_2 \vphantom{\frac{1072}{9}}\right) \ddc +  \left(-\frac{1616}{27} + \frac{176\zeta_2}{3}  + 56 \zeta_3 \right) \ddb +  \left( -\frac{1535}{12} + \frac{592\zeta_2}{9} + 28 \zeta_3 \right.
\nn \\ &
\left. - 6 \zeta_4 \vphantom{\frac{1535}{12}}\right) \dda \bigg] C_A C_F +  \bigg[ \frac{32}{3} \ddd - \frac{160}{9} \ddc + \left( \frac{224}{27} - \frac{32\zeta_2}{3}  \right) \ddb +  \left(\frac{127}{6} - \frac{112\zeta_2}{9} \right.
\nn \\ &
\left. + 8\zeta_3 \vphantom{\frac{127}{6}}\right) \dda \bigg] C_F N_f \bigg\}+ a^3 \bigg\{ \bigg[ 512 \ddg +  \Big( -2048 - 3072\zeta_2 \Big)\dde + 10240 \zeta_3 \ddd  + \Big( 2044
\nn \\ &
+ 2976\zeta_2 - 960\zeta_3 - 7104 \zeta_4 \Big) \ddc+ \Big( -4096 \zeta_3 - 6144 \zeta_2 \zeta_3 +12288 \zeta_5 \Big) \ddb + \left( -\frac{5599}{6} \right.
\nn \\ &
\left. - \frac{130}{3}\zeta_2 - 460\zeta_3 + 206 \zeta_4 + 80 \zeta_2 \zeta_3 + 1328 \zeta_5 + \frac{10336}{3} \zeta_3^2 - \frac{23092}{9}\zeta_6 \vphantom{\frac{5599}{6}}\right) \dda \bigg] C_F^3 
\nn \\ &
+ \bigg[ - \frac{7040}{9} \ddf + \left( \frac{17152}{9} - 512 \zeta_2  \vphantom{\frac{17152}{9}}\right) \dde + \left( -\frac{4480}{9} + \frac{11264\zeta_2}{3} + 1344\zeta_3 \right) \ddd 
\nn \\ &
+ \left(-\frac{35572}{9} - \frac{11648}{9} \zeta_2 - 5184 \zeta_3 + 1824 \zeta_4 \right) \ddc + \left( \frac{25856}{27} - \frac{12416\zeta_2}{27} + \frac{26240\zeta_3}{9}\right.
\nn \\ &
\left.  + \frac{3520\zeta_4}{3} - 1472 \zeta_2 \zeta_3 \right) \ddb + \left( \frac{74321}{36} - \frac{13186\zeta_2}{27} - \frac{20156\zeta_3}{9} - \frac{832\zeta_4}{27} + \frac{28736\zeta_2 \zeta_3}{9} \right. 
\nn \\ &
\left.  - \frac{39304\zeta_5}{9} + \frac{3280\zeta_3^2}{3} - \frac{2602\zeta_6}{9}  \right) \dda \bigg] C_A C_F^2 + \bigg[ \frac{7744}{27} \dde + \left( -\frac{28480}{27} + \frac{704\zeta_2}{3} \right) \ddd
\nn \\ & 
+ \left(  \frac{124024}{81} - \frac{12032\zeta_2}{9}  - 704 \zeta_3 + 352 \zeta_4 \vphantom{\frac{124024}{81}}\right) \ddc +  \left( -\frac{594058}{729} + \frac{98224\zeta_2}{81} + \frac{40144\zeta_3}{27}  \right.
\nn \\ &
\left. - \frac{1496\zeta_4}{3} - \frac{352\zeta_2 \zeta_3}{3} - 384\zeta_5 \vphantom{\frac{594058}{729}}\right) \ddb +  \left( -\frac{1505881}{972} + 843 \zeta_2 + \frac{82385\zeta_3}{81} + \frac{14611\zeta_4}{54}  \right.
\nn \\ & 
\left. - \frac{884\zeta_2 \zeta_3}{3} - 204 \zeta_5 - \frac{400\zeta_3^2}{3} + \frac{1658\zeta_6}{9} \right)\dda \bigg] C_A^2 C_F + \bigg[\frac{1280}{9} \ddf - \frac{2560}{9} \dde 
\nn \\ &
+  \left( \frac{544}{9} - \frac{2048\zeta_2}{3} \right) \ddd  + \left( \frac{4288}{9} + \frac{2048\zeta_2}{9} + 1280\zeta_3 \right) \ddc + \left(\vphantom{\frac{736\zeta_4}{3}} -6 + \frac{1952\zeta_2}{27} \right. 
\nn \\ &
\left.- \frac{5728\zeta_3}{9} - \frac{736\zeta_4}{3} \right) \ddb+ \left( -\frac{421}{3} + \frac{2632\zeta_2}{27} + \frac{3512\zeta_3}{9} + \frac{136\zeta_4}{27} - \frac{5504\zeta_2 \zeta_3}{9} \right.
\nn \\ &
\left. + \frac{5536\zeta_5}{9} \right) \dda \bigg]C_F^2 N_f + \bigg[ - \frac{2816}{27} \dde + \left( \frac{9248}{27} - \frac{128\zeta_2}{3}  \right) \ddd + \left( -\frac{32816}{81}  \right.
\nn \\ &
\left. + 384\zeta_2  \vphantom{\frac{32816}{81}}\right) \ddc +  \left( \frac{125252}{729}- \frac{29392\zeta_2}{81} -\frac{2480\zeta_3}{9} + \frac{368\zeta_4}{3} \right) \ddb + \left( \frac{110651}{243}  \right.
\nn \\ &
\left. - \frac{28132\zeta_2}{81} - \frac{6016\zeta_3}{81} - \frac{2878\zeta_4}{27} + \frac{208\zeta_2 \zeta_3}{3}  - 8 \zeta_5 \right) \dda \bigg]  C_A C_F N_f 
\nn \\ &
 + \bigg[ \frac{256}{27} \dde - \frac{640}{27} \ddd + \left( \frac{1600}{81} - \frac{256\zeta_2}{9} \right) \ddc + \left( -\frac{3712}{729} + \frac{640\zeta_2}{27} + \frac{320\zeta_3}{27} \right) \ddb 
\nn \\ &
+  \left( -\frac{7081}{243} + \frac{2416\zeta_2}{81} - \frac{1264\zeta_3}{81}  + \frac{320\zeta_4}{27} \right) \dda \bigg] C_F N_f^2 
\nn \\ &
  + \bigg[ \left( 4 + 10 \zeta _2+\frac{14 \zeta _3}{3} - \zeta _4 - \frac{80 \zeta _5}{3} \right) \dda \bigg] \frac{d_{abc}d_{abc}}{\Nc} N_{q\gamma} \bigg\}\,,
\label{eq:DYsv}
\end{align}
where $N_{q\gamma} =  (1/e_q){\sum_{q^\prime} e_{q^\prime}}$ is the charge-weighted sum of the $N_f$ quark flavors normalized to the charge of the primary quark $q$
and $d_{abc}d_{abc} = (\Nc^2 - 1)(\Nc^2 - 4)/\Nc$.
Our calculation agrees completely both with the well-known tower of plus distributions~\cite{hep-ph/0508265,hep-ph/0508284,hep-ph/0603041,hep-ph/0605068} and with the delta function terms predicted by
two groups independently using an approach different than the one described above~\cite{1404.0366,1405.4827}.

\section{A \n3lo soft-virtual term for the $\Neqfour$ model}
\label{sec:SYM}
Loop corrections in $\Neqfour$ super Yang-Mills theory often have a simple structure and it is therefore interesting to consider an analog of the soft-virtual terms treated above in such a model.
In fact, it is natural to consider the soft-virtual corrections in a $SU(\Nc)$ $\Neqfour$ super Yang-Mills theory for the simple case of scalar pair production because the relevant hard function has already been
computed through to three-loop order~\cite{1112.4524,vanNeerven:1985ja}. As usual, $\Neqfour$ super Yang-Mills theory refers to a theory with a single massless
$\Neqfour$ supermultiplet in the adjoint representation of $SU(N_c)$. The massless $\Neqfour$ supermultiplet contains a massless gauge field, four massless Majorana fermions, and three massless complex scalars.
It therefore follows that, to obtain the desired result in the $\Neqfour$ model, we need to calculate diagrams with scalars in the loops or in the final state and appropriately adjust the color coefficients for the fermionic contributions.
No additional master integrals are required beyond those which we have already calculated. 
It is worth emphasizing that it is inconvenient to use the CDR scheme for this calculation because it does not preserve supersymmetry. 
Instead, we employ the FDH scheme~\cite{hep-ph/0202271} in which the diagrammatic numerator algebra is performed in $D_s = 4$ dimensions, while the required integral reductions are carried out in $D = 4 - 2 \e$ dimensions. 
As is well-known, the beta function vanishes in all $\Neqfour$ models to all orders in perturbation theory~\cite{Sohnius:1981sn,Howe:1983wj,Howe:1983sr,Brink:1982wv,Brink:1982pd,Lemes:2001vf}. Among other things, this implies that
no additional renormalization related to so-called $\e$-scalar contributions is necessary~\cite{1404.2171}. 

Following the formalism first outlined in reference~\cite{vanNeerven:1985ja}, we consider the annihilation of two scalars $\phi^{a\dagger}_r$ and $\phi^a_t$ into a color-singlet scalar current $J_{rt}$ plus anything,
\begin{align}
  \phi^{a\dagger}_r + \phi^a_t \to J_{rt} + X \,,
\end{align}
where the color-singlet scalar current reads
\begin{align}
  J_{rt} = \frac{1}{2} ( \phi^{a\dagger}_r \phi^a_t + \phi^{a\dagger}_t \phi^a_r ) - \frac{1}{3} \delta_{rt} \phi^{a\dagger}_s \phi^a_s \,. 
\label{eq:scalarcurrent}
\end{align}
Along the lines described above, we compute the relevant three-loop soft function and see that,
given an appropriate definition of transcendentality weight for the distributions that appear in the result, the soft function obeys a so-called maximal transcendentality weight principle.

In order to discuss the observed correspondence, let us first define the notion of transcendentality weight for the relevant terms.
As is well-known, the transcendentality weight is $0$ for rational numbers, $1$ for a logarithm, and $i$ for $\zeta_i$.
The weight of a product of these quantities is the sum of the weights of the individual factors in the product.
Taking into account the integration to be performed over $z$, we associate to the distributions introduced in Eqs. (\ref{eq:dists}), $\mathcal{D}_i$, the transcendentality weight $i$.
Given this definition, the three-loop soft function for scalar pair production in $SU(\Nc)$ $\Neqfour$ super Yang-Mills coincides
with the part of the three-loop Higgs boson production soft function which is of maximal transcendentality weight (six in this case). 
In fact, each of the contributions enumerated at the beginning of Section \ref{sec:softHiggs} separately satisfies this maximal transcendentality weight principle, similar and yet distinct from the original principle 
observed long ago in the context of the anomalous dimensions of $\Neqfour$ twist-two operators~\cite{hep-th/0404092}.
In the $\Neqfour$ theory, similar observations have subsequently been made for form factors~\cite{1112.4524,1201.4170,1203.0454} and for the tower of plus distributions in the soft-virtual term~\cite{hep-ph/0508265}.
Our analysis extends these observations to the complete set of soft-virtual corrections at \n3lo.
Combining all of the relevant ingredients together, we find
\begin{align}
  G^{\Neqfour}(z, 0) = &\dda + a \bigg\{ \bigg[ 16 \ddc + 8 \zeta_2 \dda\bigg] C_A \bigg\} + a^2 \bigg\{ \bigg[ 128 \dde - 160 \zeta_2 \ddc + 312 \zeta_3 \ddb - 2 \zeta_4 \dda \bigg] C_A^2\bigg\} 
\nn \\ &
+ a^3 \bigg\{ \bigg[ 512 \ddg - 3584 \zeta_2 \dde + 11584 \zeta_3 \ddd - 4928 \zeta_4 \ddc
\nn \\ &
 + \left(- \frac{23200 \zeta_2 \zeta_3}{3} + 11904 \zeta_5 \right) \ddb  + \left( \frac{13216 \zeta_3^2}{3} - \frac{8012 \zeta_6}{3} \right) \dda \bigg] C_A^3\bigg\}
\label{eq:Neqfour}
\end{align}
for the soft-virtual corrections to scalar pair production in $SU(\Nc)$ $\Neqfour$ super Yang-Mills theory.
One can explicitly check by comparing Eqs.~(\ref{eq:Hsv}) and~(\ref{eq:Neqfour}) that the principle of maximal transcendentality weight is satisfied.

\section{Summary and Outlook}
\label{sec:conclude}

In this work, we presented the \n3lo soft-virtual corrections to both gluon-fusion Higgs boson production and Drell-Yan lepton production in QCD and to color-singlet scalar pair production in $SU(\Nc)$ $\Neqfour$ super Yang-Mills theory. 
Our main results are given in Eqs.~(\ref{eq:delc3}), (\ref{eq:Hsv}), (\ref{eq:DYsv}), and (\ref{eq:Neqfour}). 
In particular, we found results in full agreement with a calculation of the Higgs soft-virtual term performed recently by a different group~\cite{1403.4616}. 
By performing a Casimir scaling of the Higgs result, we reproduced the \n3lo Drell-Yan soft-virtual term derived in earlier work~\cite{1404.0366,1405.4827}.
Finally, we observed a maximal transcendentality weight principle whereby one can predict the complete result for color-singlet scalar pair production in the $\Neqfour$
theory from the part of the soft-virtual term for Higgs production which is of maximal transcendentality weight.

The \n3lo Higgs soft-virtual term presented in this paper provides an important reference point for the full fixed-order calculation.
In light of recent work on the subject~\cite{1411.3584,1405.4827,1405.5685,1408.6277,1405.3654}, it is not clear that N$^3$LL soft-gluon resummation
alone will improve the prediction of QCD perturbation theory to the required extent. 
Therefore,  a full three-loop, fixed-order calculation of the partonic cross section is highly desirable 
and should be carried out in the near future in order to obtain a full N$^3$LO+N$^3$LL QCD prediction for the gluon-fusion Higgs boson production cross section.


\vskip0.5cm
\noindent {\large\bf Acknowledgments}
\vskip0.3cm
\noindent YL and HXZ thank Lance~Dixon, Michael~Peskin, and Stefan~H\"{o}che for useful discussions. HXZ thanks Adrian~Signer for insightful comments on the use of the FDH scheme. 
YL and HXZ would also like to thank the Institute of High Energy Physics in China for hospitality while part of this work was carried out.
Our figure was generated using {\tt Jaxodraw}~\cite{hep-ph/0309015}, based on {\tt AxoDraw}~\cite{CPHCB.83.45}.
The research of YL and HXZ is supported by the US Department of Energy under contract DE-AC02-76SF00515. 
The research of RMS is supported in part by the ERC Advanced Grant EFT4LHC of the European Research Council, the Cluster of Excellence Precision Physics, Fundamental Interactions and Structure of Matter (PRISMA-EXC 1098).

\appendix

\section{Anomalous dimensions and beta function coefficients}
\label{sec:append}
In this appendix, we present the anomalous dimensions and beta function coefficients relevant to the calculation of the terms in the Laplace-transformed soft function for gluon-fusion Higgs boson production which are fixed by RG invariance.
Through to three loops, the perturbative coefficients of the cusp anomalous dimension, $\gcusp^{\rm H}(a)$, are given by~\cite{hep-ph/0403192}
\begin{align}
  \Gamma^{\rm H}_0 = & \,4 C_A
\\ 
  \Gamma^{\rm H}_1 = & \left( \frac{268}{9} - 8 \zeta_2 \right) C_A^2 - \frac{40}{9} C_A N_f
 \\
\Gamma^{\rm H}_2 = &  \left( \frac{490}{3} - \frac{1072\zeta_2}{9}  + \frac{88\zeta_3}{3}  + 88 \zeta_4 \right) C^3_A +  \left( -\frac{836}{27} + \frac{160\zeta_2 }{9}- \frac{112\zeta_3}{3} \right)C_A^2 N_f
\nn \\ &
 +  \left( - \frac{110}{3} + 32 \zeta_3 \right)C_A C_F N_f - \frac{16}{27} C_A N_f^2\,.
\end{align}

The perturbative coefficients of the soft anomalous dimension, $\ghs^{\rm H}(a)$, can be extracted from the pole terms of the perturbative coefficients of the gluon form factor~\cite{hep-ph/0508055}.
Through to three loops, they read
\begin{align}
  \gamma_0^{\rm H} = & \,0
\\ 
\gamma_1^{\rm H} = & \left( -\frac{808}{27}  + \frac{22\zeta_2}{3} + 28 \zeta_3 \right)C_A^2 + \left( \frac{112}{27}  - \frac{4 \zeta_2}{3}  \right) C_A N_f
\\
\gamma_2^{\rm H} = & \left( - \frac{136781}{729} + \frac{12650\zeta_2}{81}  + \frac{1316\zeta_3}{3} - 176 \zeta_4 - \frac{176\zeta_2 \zeta_3}{3}  - 192 \zeta_5 \right) C_A^3
\nn
\\ &
+\left( \frac{11842}{729} - \frac{2828\zeta_2}{81}  - \frac{728\zeta_3}{27} + 48\zeta_4 \right) C_A^2 N_f + \left( \frac{1711}{27} - 4 \zeta_2 - \frac{304\zeta_3}{9}  - 16\zeta_4 \right) C_A C_F N_f
\nn 
\\ &
 + \left( \frac{2080}{729} + \frac{40\zeta_2}{27} - \frac{112\zeta_3}{27} \right) C_A N_f^2\, .
\end{align}

Finally, the first two coefficients of the QCD beta function are required. They are given by (see {\it e.g.}~\cite{hep-ph/9701390})
\begin{align}
  \beta_0 = &\frac{11}{3} C_A - \frac{2}{3} N_f
\\ 
\beta_1 = & \frac{34}{3} C_A^2 - \frac{10}{3} C_A N_f - 2 C_F N_f\, .
\end{align}

\end{document}